\documentclass[conference]{IEEEtran}
\IEEEoverridecommandlockouts
\usepackage{cite}
\usepackage{amsmath,amssymb,amsfonts}
\usepackage{algorithmic}
\usepackage{graphicx}
\usepackage{textcomp}
\usepackage{xcolor}
\def\BibTeX{{\rm B\kern-.05em{\sc i\kern-.025em b}\kern-.08em
    T\kern-.1667em\lower.7ex\hbox{E}\kern-.125emX}}
    
\usepackage[ampersand]{easylist}    
\usepackage[hyphens]{url}
\usepackage{pifont}    

\newcommand{\linebreakand}[1][\hfill]{
  \end{@IEEEauthorhalign}
  \hfill\mbox{}\par
  \mbox{}\hfill\begin{@IEEEauthorhalign}
}

\begin{document}

% \title{CrowdHub: Extending crowdsourcing platforms for the controlled evaluation of tasks designs\\
\title{Challenges and strategies for running controlled crowdsourcing experiments \\
\thanks{This work was supported by the Russian Science Foundation (Project No. 19-18-00282).
This is a post-peer-review, pre-copyedit version of an article accepted to the XLVI Latin American Computer Conference, CLEI 2020.
}
}

% \author{
% \IEEEauthorblockN{Jorge Ram\'irez}
% \IEEEauthorblockA{%\textit{Dept. of Information Engineering and Computer Science} \\
% \textit{University of Trento}\\
% Trento, Italy \\
% jorge.ramirezmedina@unitn.it}
% \and
% \IEEEauthorblockN{Marcos Baez}
% \IEEEauthorblockA{%\textit{dept. name of organization (of Aff.)} \\
% \textit{Université Claude Bernard}\\
% \textit{Lyon 1}\\
% Lyon, France \\
% marcos-antonio.baez-gonzalez@univ-lyon1.fr
% }
% \and
% \IEEEauthorblockN{Fabio Casati}
% \IEEEauthorblockA{%\textit{dept. name of organization (of Aff.)} \\
% \textit{University of Trento}\\
% Trento, Italy \\
% casati@disi.unitn.it
% }
% \linebreakand
% \IEEEauthorblockN{Luca Cernuzzi}
% \IEEEauthorblockA{%\textit{dept. name of organization (of Aff.)} \\
% \textit{Catholic University}\\
% \textit{Nuestra Señora de la Asunci\'on}\\
% Asunci\'on, Paraguay \\
% lcernuzz@uc.edu.py
% }
% \and
% \IEEEauthorblockN{Boualem Benatallah}
% \IEEEauthorblockA{%\textit{dept. name of organization (of Aff.)} \\
% \textit{University of New South Wales}\\
% Sydney, Australia \\
% boualem@cse.unsw.edu.au
% }
% }

\author{
\IEEEauthorblockN{
Jorge Ram\'irez$^{+}$, Marcos Baez$^{\Diamond}$, Fabio Casati$^{+}$, Luca Cernuzzi$^{\ast}$, and Boualem Benatallah$^{\triangle}$
}
\IEEEauthorblockA{
$^{+}$University of Trento, Italy. $^{\Diamond}$Université Claude Bernard Lyon 1, France. \\
$^{\ast}$Catholic University Nuestra Señora de la Asunci\'on, Paraguay. $^{\triangle}$University of New South Wales, Australia.
}
}

\maketitle

\begin{abstract}
This paper reports on the challenges and lessons we learned while running controlled experiments in crowdsourcing platforms. Crowdsourcing is becoming an attractive technique to engage a diverse and large pool of subjects in experimental research, allowing researchers to achieve levels of scale and completion times that would otherwise not be feasible in lab settings.  However, the scale and flexibility comes at the cost of multiple and sometimes unknown sources of bias and confounding factors that arise from  technical limitations of crowdsourcing platforms and from the challenges of running controlled experiments in the 
``wild". In this paper, we take our experience in running systematic evaluations of task design as a motivating example to explore, describe, and quantify the potential impact of running uncontrolled crowdsourcing experiments and derive possible coping strategies. Among the challenges identified, we can mention sampling bias, controlling the assignment of subjects to experimental conditions,  learning effects, and reliability of crowdsourcing results.  According to our empirical studies, the impact of potential biases and confounding factors can amount to a 38\% loss in the utility of the data collected in uncontrolled settings; and it can significantly change the outcome of experiments. These issues ultimately inspired us to implement CrowdHub, a system that sits on top of major crowdsourcing platforms and allows researchers and practitioners to run controlled crowdsourcing projects.
\end{abstract}


\section{Background \& Motivation}

% In this paper, we present CrowdHub, a system that sits on top of major crowdsourcing platforms and allows researchers and practitioners to run controlled crowdsourcing projects.

% {
% \color{blue}
% add gist sentence according to the new framing

% connect the background and motivation with "running controlled experiments"
% }

% TODO(jr): We introduce a running example and use case to illustrate the challenges that arise when we want to study multiple task design alternatives in a crowdsourcing platform.

A crucial aspect in running a successful crowdsourcing project is identifying an appropriate task design \cite{DBLP:journals/pvldb/JainSPW17}, typically consisting of trial-and-error cycles.
Task design goes beyond defining the actual task interface, involving the deployment, collection, and mechanisms for assuring the contributions meet quality objectives \cite{DBLP:journals/csur/DanielKCBA18}.

The design of a task represents a multi-dimensional challenge.
The instructions are vital for communicating the needs of requesters since poorly defined guidelines could affect the quality of the contributions \cite{DBLP:conf/hcomp/WuQ17,DBLP:conf/cscw/KitturNBGSZLH13,DBLP:conf/ht/GadirajuYB17,DBLP:conf/naacl/LiuSBLLW16}, as well as task acceptance \cite{DBLP:conf/ecis/SchulzeSGKS11}.
Moreover, enriched interfaces could also help workers in performing tasks faster, and with results of potentially higher-quality \cite{DBLP:conf/chi/SampathRI14,Wilson2016WWW,ramirez2019}.
Accurate task pricing is also a relevant aspect since workers are paid for their contributions \cite{Whiting2019FairWC}, representing an incentive mechanism that impacts the number of worker contributions  \cite{DBLP:journals/sigkdd/MasonW09}, as well as the quality, especially for demanding tasks \cite{DBLP:conf/www/HoSSV15}.
The time workers are allowed to spend on a task can also affect the quality of the contributions \cite{maddalena2016crowdsourcing,krishna2016embracing}, and even characteristics of the crowd marketplace and work environment \cite{DBLP:journals/imwut/GadirajuCGD17}. 
These insights constitute guidelines for articulating effective task designs and highlight the feasibility of running controlled experiments in crowdsourcing platforms.
%These insights constitute guidelines for articulating effective task designs. Moreover, these results come from carefully-designed and executed experiments that highlight the feasibility of running controlled experiments in crowdsourcing platforms.

\begin{figure}[t]
\centering
\includegraphics[width=\columnwidth]{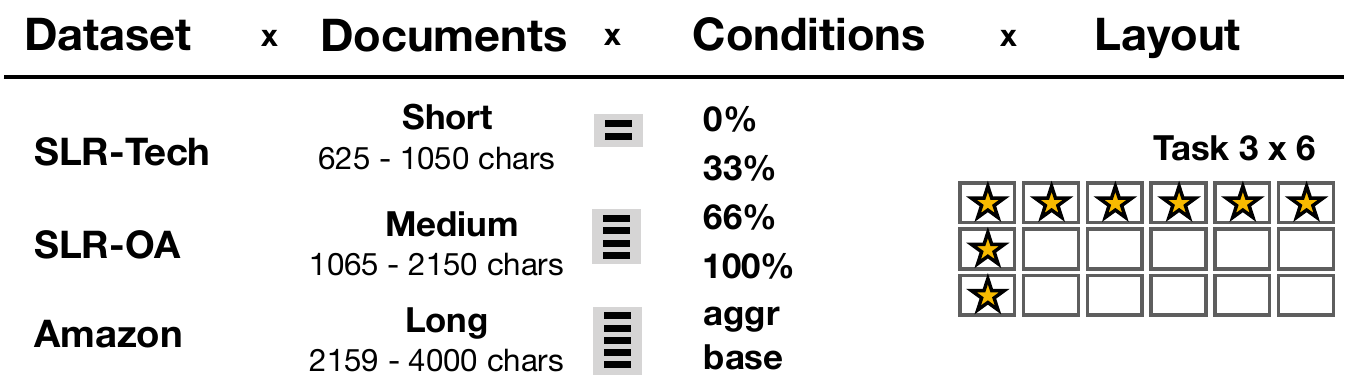}
\caption{A summary of the experimental design used to discuss the challenges of running controlled crowdsourcing experiments. We use this experiment as our running example, and its goal is to study the impact of text highlighting in crowdsourcing tasks. In this case, the experiment uses a between-subjects design and considers datasets from multiple domains with documents of varying sizes, six experimental conditions, and tasks organized in pages. The figure is adapted from \cite{ramirez2019}.}
\label{fig:highlighting-study}
\end{figure}

Over the years, a vast body of work grew the scope of crowdsourcing \cite{Paolacci2010RunningEO,Buhrmester2011AmazonsMT,DBLP:journals/sigkdd/MasonW09,Schnoebelen2010UsingAM,DBLP:conf/icse/SunS16,Crump2013}, expanding its application beyond serving as a tool to create machine learning datasets \cite{DBLP:conf/emnlp/SnowOJN08,DBLP:conf/naacl/LiuSBLLW16}. 
Paid crowd work thus establishes as a mechanism for running user studies \cite{DBLP:conf/chi/KitturCS08,Buhrmester2011AmazonsMT,DBLP:conf/icse/SunS16}, complex work that initially does not fit in the microtask market \cite{DBLP:conf/uist/KitturSKK11,DBLP:conf/uist/AhmadBMK11,DBLP:conf/cscw/KulkarniCH12}, and experiments beyond task design evaluation \cite{Crump2013,Paolacci2010RunningEO,DBLP:journals/sigkdd/MasonW09,Schnoebelen2010UsingAM}.
%\cite{Crump2013,Paolacci2010RunningEO,DBLP:journals/sigkdd/MasonW09,Schnoebelen2010UsingAM,DBLP:conf/hcomp/WuQ17,DBLP:conf/cscw/KitturNBGSZLH13,DBLP:conf/ht/GadirajuYB17,DBLP:conf/chi/SampathRI14,Wilson2016WWW,ramirez2019,maddalena2016crowdsourcing,krishna2016embracing}.
Naturally, the set of challenges also increases along with the scope and ambition of crowdsourcing projects, especially for crowdsourcing experiments.

Figure \ref{fig:highlighting-study} depicts the study we use as our running example throughout the paper to describe the challenges and strategies for running controlled experiments in crowdsourcing platforms.
The goal of this project was to understand \textit{if}, and \textit{under what conditions}, highlighting text excerpts relevant to a given relevance question would improve worker performance \cite{ramirez2019,RamirezBMC2019}. 
This required testing different highlighting conditions (of varying quality) against a baseline without highlighting, given different document sizes and datasets of different characteristics. 
The resulting experimental design featured a combination of \textit{dataset} (3) x \textit{document size} (3) x \textit{highlighting conditions} (6) --- a total of 54 configurations.

The potential size of the design space, along with the individual and environmental biases \cite{DBLP:conf/chi/BarbosaC19,DBLP:conf/lrec/BalahurSKZGHPB10,DBLP:conf/ht/ChengC13,eickhoff2018cognitive,DBLP:conf/coling/NguyenTDGTMJ14,DBLP:conf/cscw/SenGGHLNRWH15}, and the limitations of crowdsourcing platforms \cite{Qarout2019PlatformRelatedFI,DBLP:conf/www/Paritosh12}, makes it difficult to run controlled crowdsourcing experiments. 
This means that researchers need to deal with the challenging task of mapping their study designs as simple tasks, managing the recruitment and verification of subjects, controlling for the assignment of subjects to tasks, the dependency between tasks, and controlling for the different inherent biases in experimental research. This requires in-depth knowledge of experimental methods, known biases in crowdsourcing platforms, and programming using the extension mechanisms provided by crowdsourcing platforms.

Current systems that extend crowdsourcing platforms focus on specific domains \cite{DBLP:conf/uist/BernsteinLMHAKCP10,CrowdRev2018,DBLP:conf/criwg/CorreiaSPF18,DBLP:conf/sigmod/FranklinKKRX11} or kinds of problems that split into interconnected components \cite{DBLP:conf/uist/KitturSKK11,DBLP:conf/uist/AhmadBMK11,DBLP:conf/cscw/KulkarniCH12}. The general purpose tooling available to task requesters comes in the form of extensions to (or frameworks build on top of) programming languages \cite{DBLP:conf/uist/LittleCGM10,DBLP:conf/socinfo/MinderB12,DBLP:conf/oopsla/BarowyCBM12}, which could potentially demand considerable work or lock the requester to a specific crowdsourcing platform.

\textbf{Contributions.} First, this paper describes the challenges and strategies for running controlled crowdsourcing experiments, as a result of the lessons we learned while running experiments in crowdsourcing platforms.
And second, we introduce CrowdHub\footnote{\url{https://github.com/TrentoCrowdAI/crowdhub-web}}, a web-based platform for running controlled crowdsourcing projects.
CrowdHub blends the flexibility from programming with requester productivity, offering a diagramming interface to design and run crowdsourcing projects.
It offers features for systematically evaluating task design to aid researchers and practitioners during the design and deployment of crowdsourcing projects across multiple platforms, as well as features for researchers to run controlled experiments.
\section{Related work}

Crowdsourcing platforms such as Amazon Mechanical Turk, Figure Eight, or Yandex Toloka, expose low-level APIs for common features associated with publishing a task to a pool of online workers. These features involve creating a task with a given template, uploading data units, submitting and keeping track of the progress, rewarding workers, defining quality control mechanisms, among others. Naturally, exposing APIs open the room for additional extensions to the feature space, and we describe the related literature on technologies that extend the capabilities of crowdsourcing platforms. We identify that these technologies can be roughly categorized in domain-specific tooling and general-purpose platforms.

\textbf{Domain-specific tools}. Soylent \cite{DBLP:conf/uist/BernsteinLMHAKCP10} is a word processor that offers three core functionalities that allow requesters to ask crowdsourcing workers to edit, proofread, or perform an arbitrary task related to editing text documents. Soylent articulated and implemented the idea that crowdsourcing could be embedded in interactive interfaces and support requesters in solving complex tasks.
Tools to support researchers in performing systematic literature reviews (SLRs) have also been developed. 
CrowdRev \cite{CrowdRev2018} is a platform that allows users (practitioners and human computation researchers) to crowdsource the screening step of systematic reviews. Practitioners can leverage crowd workers via an easy-to-use web interface, leaving CrowdRev in charge of dealing with the intricacies of crowdsourcing. For human computation researchers, however, CrowdRev offers the flexibility to tune (and customize) the algorithms involved in the crowdsourcing process (i.e., querying and aggregation strategies).
SciCrowd \cite{DBLP:conf/criwg/CorreiaSPF18} is a system that embeds crowdsourcing workers in a continuous information extraction pipeline where human and machine learning models collaborates to extract relevant information from scientific documents.

% crowd for data cleansing tasks
Crowdsourcing databases extend the capabilities of database systems to allow answering queries via crowd workers to aid data cleansing pipelines. 
CrowdDB \cite{DBLP:conf/sigmod/FranklinKKRX11} extends the standard SQL by introducing crowd-specific operators and extensions to the data definition language (DDL) that the query engine can interpret and spawn crowdsourcing tasks accordingly. CrowdDB manages the details related to publishing crowdsourcing tasks, automatically generating task interfaces based on the metadata specified by developers using the extended DDL.
Several other declarative approaches were proposed to embed crowdsourcing capabilities into query processing systems \cite{DBLP:conf/cikm/ParameswaranPGPW12,DBLP:conf/cidr/DemartiniTKF13,DBLP:journals/pvldb/MarcusWKMM11,DBLP:journals/pvldb/MorishimaSMAF12}.

\textbf{General-purpose platforms}. Several approaches have been proposed to manage complex problems that partition into interdependent tasks. Inspired by the MapReduce programming paradigm \cite{DBLP:conf/osdi/DeanG04}, CrowdForge \cite{DBLP:conf/uist/KitturSKK11} offers a framework to allow solving complex problems via a combination of partition, map and reduce tasks.
Jabberwocky \cite{DBLP:conf/uist/AhmadBMK11} is a social computing stack that offers three core components to tackle complex (and potentially interdependent) tasks. Dormouse represents the foundations of Jabberwocky, and it acts as the runtime that can process human (and hybrid) computation, offering for cross-platform capabilities (e.g., going beyond Amazon Mechanical Turk). The ManReduce layer is similar to CrowdForge but implemented as a framework for the Ruby programming language, allowing map and reduce steps to be executed by either crowd or machines. Finally, Jabberwocky offers a high-level procedural language called Dog that compiles down to ManReduce programs. 
Turkomatic \cite{DBLP:conf/cscw/KulkarniCH12}, unlike previous approaches, allows the crowd to play an active role in decomposing the problem into the set of interdependent components. Turkomatic operates using a divide-and-conquer loop where workers actively refine the input problem (supervised by requesters) into subtasks that run on AMT, where a set of generic task templates are instantiated accordingly.

\begin{figure*}[t]
\centering
\includegraphics[width=0.8\textwidth]{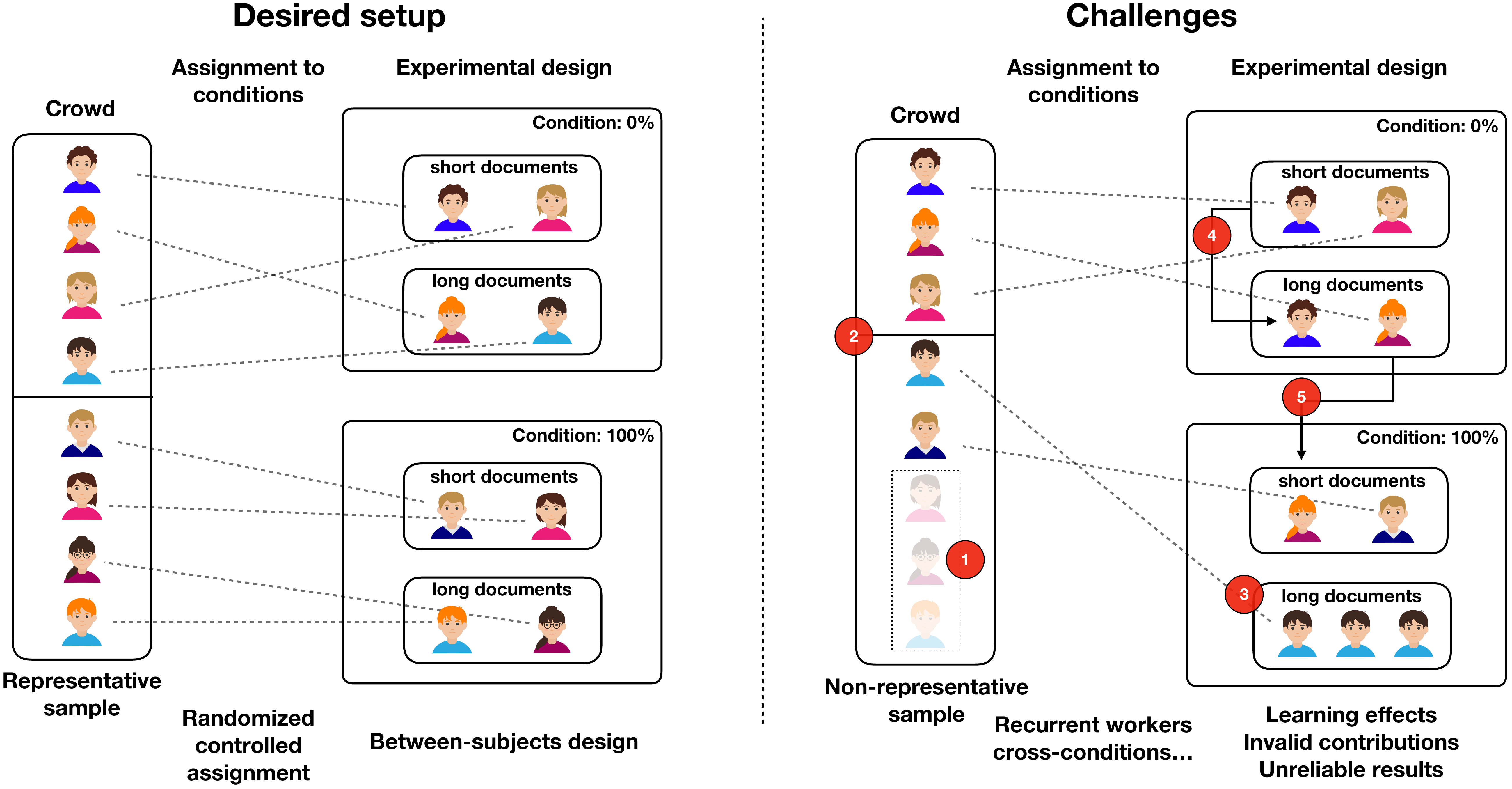}
\caption{ The challenges that arise while running controlled crowdsourcing experiments. }
\vspace{-10pt}
\label{fig:challenges}
\end{figure*}

Turkit \cite{DBLP:conf/uist/LittleCGM10} is a JavaScript programming toolkit for implementing human computation algorithms that run on Amazon Mechanical Turk. It offers functions that interface with AMT and introduce the crash-and-rerun programming model for building robust crowdsourcing scripts.
CrowdLang \cite{DBLP:conf/socinfo/MinderB12} is a framework and programming language for human and machine computation, relying on three core components to allow requesters to implement complex crowdsourcing workflows. First, a programming library that encapsulates operators and reusable computation workflows (e.g., find-fix-verify \cite{DBLP:conf/uist/BernsteinLMHAKCP10}, iterative improvements \cite{DBLP:conf/uist/LittleCGM10}). Then, the engine component orchestrates and run human computation algorithms (and deals with the underlying challenges). And last, the integration layer bridges CrowdLang with distinct crowdsourcing platforms for cross-platform support.
Similar to Turkit, AUTOMAN \cite{DBLP:conf/oopsla/BarowyCBM12} embeds crowdsourcing capabilities into a programming language (Scala in this case). However, it also offers features to automatically manage pricing, quality control, and scheduling of crowdsourcing tasks.

Substantial efforts have been devoted to offering solutions that sit on top of crowdsourcing platforms. However, these tools still demand considerable work from requesters (e.g., programming the complete workflows or experiments).
For running controlled experiments, we find the TurkServer platform \cite{DBLP:conf/hcomp/MaoCGPPZ12} to be closely related to our work. TurkServer is based on a JavaScript web framework and offers builtin features that enable researchers to run experiments on AMT.
Even though programming frameworks give full flexibility to researchers, CrowdHub aims to offer a paradigm that blends flexibility with productivity. 
Concretely, we propose an easy-to-use platform that allows task requesters to focus on designing crowdsourcing workflows via a diagramming interface, reducing the programming efforts that current tooling would require.

\section{Challenges in Evaluating Task Designs}

% after adding the running example in the Background, this section goes directly into mentioning the challenges.

% {
% \color{blue}
% 1) timezones
% 2) controlling demographics
% 3) a subgroup of workers (e.g., a country) could dominate the task
% 4) recurrent workers
% 5) condition crossover
% }

In this section, we describe the challenges that arise when running crowdsourcing experiments. These challenges are derived from our own experiences in evaluating task designs while studying the impact of highlighting support in text classification \cite{ramirez2019,RamirezBMC2019}. 

We return to the study we use as our running example (see Figure \ref{fig:highlighting-study}) to describe the challenges while running controlled crowdsourcing experiments.
The datasets used in this experiment come from two domains: systematic literature reviews (SLR) and Amazon product reviews. Using the number of characters as a proxy for document length, we categorized the documents as short, medium, or long. 
The workers that participated in the experiment performed the task in pages (up to six), where each page showed three documents with one item used for quality control\cite{DBLP:journals/csur/DanielKCBA18}, except the first page which was used entirely for quality control. 
Each document in the datasets was associated with a list of text excerpts of varying quality. The highlighting conditions 0\%-100\% indicate the proportion of items in a given page that will highlight text excerpts of good quality (0\% means non-useful highlights). The baseline was used as the control condition (no highlights), and the \textit{aggr} condition first aggregated the available excerpts and then highlighted the result.

The left half of Figure \ref{fig:challenges} depicts the desired experimental setup, a between-subjects design. A representative sample of workers from the selected crowdsourcing platform is randomly assigned to one of the experimental conditions, assuring a well-balanced distribution of participants. Also, within each of the conditions, workers are restricted to assess only documents of a specific size.
However, running this experiment on crowdsourcing platforms is far from being a straightforward task. Researchers need to put a lot of effort into overcoming the limitations of crowdsourcing platforms and successfully executing the desired experimental design. And in this process, challenges emerge that could hurt the outcome and validity of the experiment, as depicted in the right half of Figure \ref{fig:challenges}.

In order to identify the challenges and quantify potential experimental biases in running an \textit{uncontrolled} evaluation of task designs, we created individual tasks in Figure Eight for a subset (1 dataset) of the experimental conditions. We ran the tasks one after another, collecting a total of $6993$ votes from $631$ workers (16 tasks). In the following, we lay out the challenges that we encountered during the process (Figure \ref{fig:challenges}).

\textbf{Platforms lack native support for experiments.} Crowdsourcing platforms such as Figure Eight (F8), Amazon Mechanical Turk (AMT), and Toloka offer the building blocks to design and run crowdsourcing tasks. In F8, for example, this implies defining i) data units to classify, ii) gold data to use for quality control, iii) task design, including instructions, data to collect, assignment of units to workers, iv) the target population (country, channels, trust), and v) the cost per worker contribution. 
F8 then manages the assignment of data units, the data collection and the computation of basic completion metrics. 
These features are suitable for running individual tasks, but less so when experimenting with different task designs with a limited pool of workers, where special care must be taken to run even simple between-subject designs \cite{DBLP:conf/chi/KitturCS08}.

This lack of support left researchers with the laborious job of actually implementing the necessary mechanisms to deploy a controlled experiment.
For our running example, this means that researchers need to create the tasks for each of the experimental conditions and document sizes ($6$ conditions and $3$ document sizes, a total of $18$ tasks). Eligibility control mechanisms are crucial to identify workers and randomly assign them to one of the tasks, controlling that workers only participate once.
Besides, during deployment, researchers need to constantly monitor the progress of the crowdsourcing task to avoid potential demographic biases presence in the crowdsourcing platforms, assuring that a well-balanced and representative sample of the population participates in the experiment.
Ultimately, this means that researchers need deep knowledge in both programming and experimental methods to implement the necessary mechanisms and controlling inherent biases in experimental research and crowdsourcing platforms.

\textbf{Timezones \ding{202}}. A wide range of countries constitutes the population of workers in crowdsourcing platforms. However, the majority of workers tend to come from a handful of countries instead.
The pool of workers that can participate in a task varies at different times of the day since workers come from a diverse set of countries with different timezones.
For example, the population of active US workers could be at its pick while workers from India are just starting the day, as shown by Difallah, Filatova, and Ipeirotis for the AMT platform \cite{DBLP:conf/wsdm/DifallahFI18}.

Running experiments without considering the mismatch of worker availability during the day could introduce confounding factors that hurt the experiment's results. In the case of our running example, this means that the experimental conditions may not be comparable.
For instance, we noticed the worker performance in independent runs of our study varied by different factors even between runs of the same condition (e.g., from $24$s to $14$s in decision time between a first and a second run considering only new workers).

Collecting reliable and comparable results thus requires multiple systematic runs over an extended period of time. For our study, this means executing the experiment over chunks of time during the day and spread it over weeks, balancing across the experimental conditions.

\textbf{Population demographics \ding{203} \ding{204}}. The demographics of a crowdsourcing platform (e.g., gender, age, country) defines the pool of active workers, along with the time an experiment runs.
Researchers tend to resort to crowdsourcing since it represents a mechanism to access a large pool of participants. However, demographic variables tend to follow a heavy-tailed distribution. For example, workers from the US and India constitute the majority of the available workforce in Amazon Mechanical Turk \cite{DBLP:conf/wsdm/DifallahFI18}.
Many kinds of experiments could potentially be sensitive to the underlying population demographics. Thus, ensuring a diverse set of workers is a crucial endeavor that researchers must undertake to perform methodologically sound experiments.

Uncontrolled worker demographics could result in an imbalanced sample of the population \ding{203} and subgroups of workers dominating the task \ding{204}, potentially amplifying human biases and produce undesired results \cite{DBLP:conf/chi/BarbosaC19}.
In running uncontrolled tasks, we observed a participation dominated by certain countries, which prevented more diverse population characteristics. For example, the top contributing countries provided 48.1\% of the total judgements (Venezuela: $28.5$\%, Egypt: $11.8$\%, Ukraine: $7.8$\%).

Crowdsourcing platforms offer basic demographic variables that can be tuned to control the population of workers participating in tasks. For our experiment, we identified the top three contributing countries and created buckets with each of the top countries as head of the groups. We then manually assigned the country buckets to the experimental conditions, distributing uniformly the tasks that these could perform and swapping buckets accordingly. This tedious but effective mechanism allowed us to overcome the heavy-tail distribution of the population demographics and give equal opportunity to the top countries that constitute the crowdsourcing platform.

\textbf{Recurrent workers may impact the results \ding{205}}. While returning workers are desirable in any crowdsourcing project, they represent a potential source of bias in the context of crowdsourcing experiments, and special care must be taken to obtain independent contributions within and across similar experiments \cite{Paolacci2010RunningEO}.

\begin{figure}[h]
\centering
\includegraphics[width=\columnwidth]{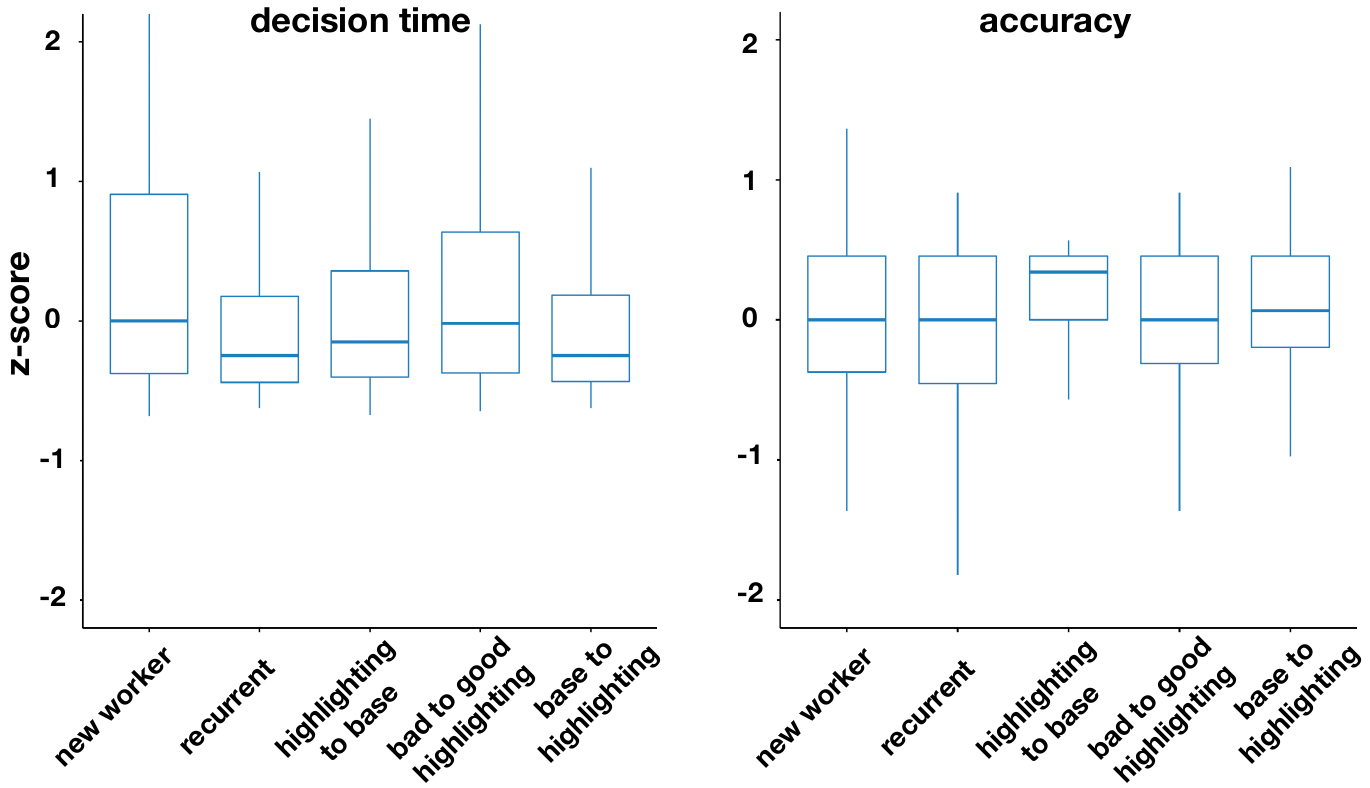}
\caption{Decision time and accuracy for recurrent workers in the highlighting support experiment. For comparison purposes, the performance values from all conditions are aggregated using a normalized z-score that considers the median from the valid contributions in its computation. The distribution of values from non-valid contributions, organized under the different sources of bias, thus depict the deviation in performance from the normal population (i.e., new workers).}
\vspace{-10pt}
\label{fig:recurrent}
\end{figure}

In our study, we published the combination of experimental conditions and document sizes as independent tasks in the crowdsourcing platform. Therefore, since tasks run in parallel, nothing prevents workers from proceeding with another of our tasks upon finishing the task where they first landed, which is the scenario depicted by point \ding{205} in Figure \ref{fig:challenges}.
This situation is potentially problematic since workers that return to complete more tasks might perform better due to the \textit{learning effect}.
As shown in Figure \ref{fig:recurrent}, we observed 38\% of returning workers, who featured a lower completion time (i.e., workers were faster) but not higher accuracy\footnote{We noticed, however, that accuracy remained mostly unaffected by conditions and other factors across all our experiments, and it might have been less susceptible to the learning effect.}.

Researchers must implement custom eligibility control mechanisms to deal with recurrent workers, and in general, to manage the population of workers that participate in the experimental conditions.
Fortunately, crowdsourcing platforms provide the necessary means to extend its set of features. The task interface shown to workers is usually a combination of HTML, JavaScript, and CSS that researchers need to code. As part of this interface, special logic could be embedded to control worker participation.
For our study, we identified workers by levering browser fingerprinting \cite{gadiraju2017improving} and sending this information to an external server that performed control and random assignment of workers to conditions.
Our task interface included JavaScript code that upon page load requested the server information about the worker, resulting in a ``block" or ``proceed" action that prevented or allowed the worker to continue with the task (in the case of the former the page showed a message with the reasons of the block).

\textbf{Recurrent workers may cross conditions \ding{206}}. Closely related to the previous issue is the fact that returning workers can also land in a different experimental condition, as depicted by point \ding{206} in Figure \ref{fig:challenges}. 
This scenario could make the experimental conditions difficult to compare, threatening the validity of the experiment, since returning workers that cross conditions could modify their behavior and resulting performance.

In our study, by comparing the 30\% workers who crossed the experimental conditions with the ``new workers" (those that never performed the task), we observed that switching between highlighting support and not support resulted in lower decision time (``highlighting to base" and ``base to highlighting" in Figure \ref{fig:recurrent}). 
However, those workers that came from the ``bad highlighting" condition and arrived at the condition with good highlighting support showed a higher decision time, possibly due to the lack of trust in the support.
Workers switching from support to no support also featured higher accuracy than the new workers and those returning to the same condition.

The same eligibility control mechanism via browser fingerprinting takes care of recurring workers that land in different conditions since it allows controlling for recurrent workers in the first place.

\bigskip
The above challenges emphasize that running a systematic comparison of task designs using the native building blocks of a crowdsourcing platform is thus a complex activity, susceptible to different types of experimental biases, which are costly to clean up (e.g., discarding 38\% of the contributions).
While our example may represent an extreme case, and it focuses on task design evaluation, it is still indicative of many of the challenges that task designers and researchers face in general when running controlled experiments in crowdsourcing platforms.
\section{CrowdHub Platform}

The above challenges motivated us to design and build a system that extends the capabilities of crowdsourcing platforms and allow researchers and practitioners to run controlled crowdsourcing projects. 
By extending major crowdsourcing platforms, CrowdHub offers cross-platform capabilities and aims to provide the building blocks to design and run crowdsourcing workflows, and for researchers, in particular, the features to run controlled crowdsourcing experiments.

\subsection{Design goals}
We now describe the design goals behind CrowdHub:

\smallskip
\begin{easylist}
\ListProperties(Hide=100, Hang=true,Style*=--~)
& \textbf{Offer cross-platform support}. CrowdHub extends crowdsourcing platforms and integrates the differences between these into features that allow task requesters to design tasks that can run across multiple platforms. This means that we can design a task (e.g., one that asks workers to classify images) and then run it on multiple platforms without dealing with the underlying details that set the platforms apart. 
This goal is particularly relevant for crowdsourcing experiments since there is evidence suggesting that results could vary across crowdsourcing platforms \cite{Qarout2019PlatformRelatedFI}.

& \textbf{Blend easy-of-use with flexibility}. While offering crowdsourcing capabilities by extending a programming language gives complete control to task requesters, it also demands more effort since requesters would need to code their solutions. 
With CrowdHub, we aim to mix the flexibility of programming with productivity, and thus instead offer a diagramming interface that does not require task requesters to code every piece of the crowdsourcing puzzle.

& \textbf{Support interweaving human and machine computation}. CrowdHub is designed to allow researchers and practitioners to extend its set of features and incorporate machine computation alongside human workers. This means, following on the image annotation example, that researchers could incorporate a machine learning model for classifying "easy" images and then derive "hard ones" to crowd workers.

\end{easylist}

\subsection{Architecture}

\begin{figure}[t]
\centering
\includegraphics[width=\columnwidth]{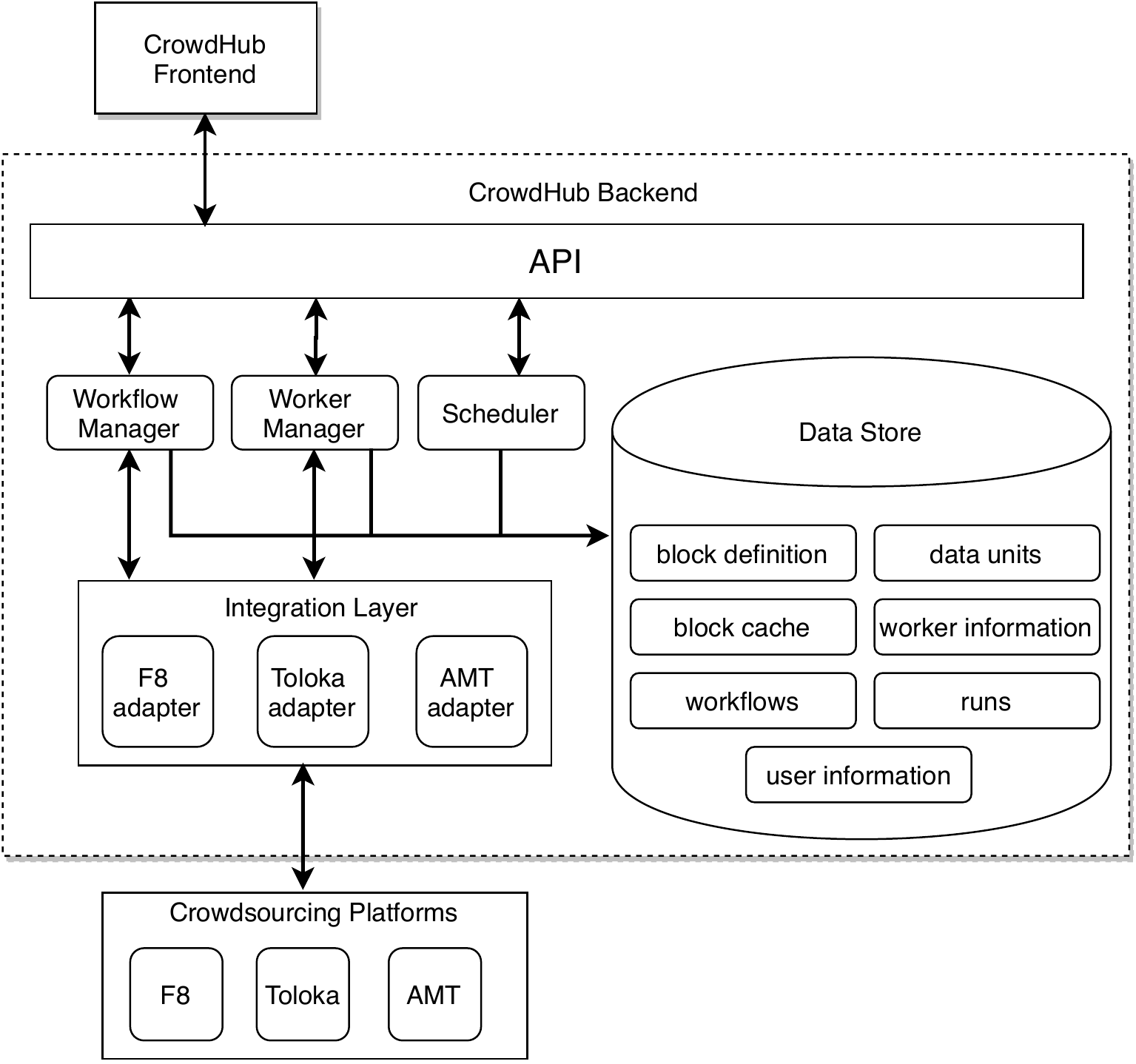}
\caption{ The architecture of the CrowdHub platform.}
\label{fig:architecture}
\end{figure}

Figure \ref{fig:architecture} shows the internal architecture of the CrowdHub platform. 
We used a client-server architecture to implement CrowdHub, where both the backend and the frontend are implemented using the JavaScript programming language.
CrowdHub offers a diagramming interface where \textit{visual blocks} are the foundation to design crowdsourcing workflows. 
A workflow is essentially a graph representing a crowdsourcing project (e.g., an experiment) that allows data units to flow through the nodes, where the nodes represent tasks or executable code to transform data units.

The \textit{\textbf{Workflow Manager}} exposes features that allow requesters to create, update, delete, and execute workflows. It uses the crash-and-rerun model \cite{DBLP:conf/uist/LittleCGM10} to offer a robust mechanism for executing workflows. 
A node in the graph defining the workflow is called \textit{block}. These blocks can be seen as ``functions" that the workflow manager can run.
The current implementation of CrowdHub offers two blocks \textit{Do} and \textit{Lambda}. 
The \textit{Do} block represents a task that is published on a crowdsourcing platform. Using a \textit{Do} block, requesters can configure the task interface using a builtin set of \textit{UI elements} that safes requesters from coding the interface (which typically consists of HTML, CSS, and JavaScript). In addition to configuring the interface,  other crowdsourcing task parameters can be specified (e.g., number of votes, monetary rewards).
The \textit{Lambda} block, accepts JavaScript code, and it represents an arbitrary function that receives and returns data (useful for data aggregation and partitioning, for example).

The \textit{\textbf{Worker Manager}} offer features for eligibility control and population management. The \textit{eligibility control} gives researcher the functionality to define the policy regarding returning workers and condition crossovers associated with the experimental design (between- or within-groups design).
Through \textit{population management} requesters can control for subgroups of workers dominating a dataset by assigning a specific quota. Altogether, these features allow requesters to be in control of their crowdsourcing project. 

\begin{figure*}[t]
\centering
\includegraphics[width=\textwidth]{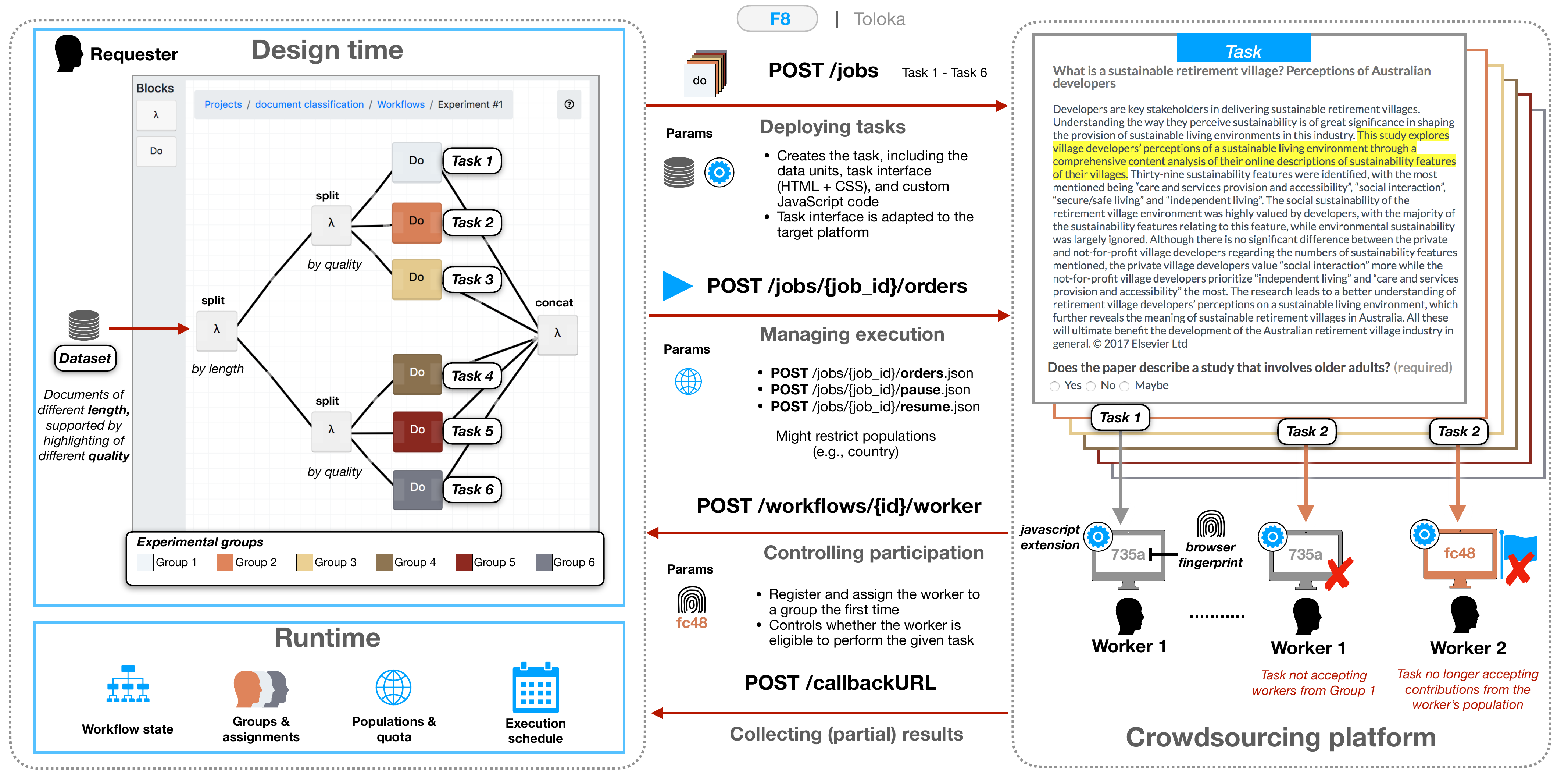}
\vspace{-15pt}
\caption{ Example workflow for a between-subjects design using CrowdHub. }
\vspace{-10pt}
\label{fig:flow}
\end{figure*}

The \textit{\textbf{Scheduler}} enables to control for confounding factors by scheduling task execution over a period of time. Task requesters specify the time and progress intervals at which the workflow should run, and the scheduler notifies the \textit{Workflow Manager} accordingly to pause and resume the execution.

The \textit{\textbf{Integration Layer}} implements the cross-platform capabilities that CrowdHub provides. This layer offers a set of functions that handle the differences between the crowdsourcing platforms, thus allowing requesters to publish tasks across multiple platforms.
The workflow manager module uses the Integration Layer to publish, pause, and resume jobs in the crowdsourcing platforms supported by CrowdHub. When publishing a task, the Integration Layer first translates the UI components that constitute the task interface into the actual interface that will be shown to workers on the selected crowdsourcing platform.
The worker manager module relies on the Integration Layer to handle worker demographics in a platform-agnostic manner, as well as implementing the eligibility control policy.
CrowdHub manages the interactions with the crowdsourcing platform through their public APIs, and JavaScript extensions incorporated into the tasks interface allows for additional features such as worker control and identification (browser fingerprinting) \cite{gadiraju2017improving}. 
The current implementation of CrowdHub supports Figure Eight and Toloka, with Amazon Mechanical Turk as a work in progress.

The \textit{\textbf{Data Store}} is a SQL database that contains user information, workflow definition and runs (which allow for running a workflow multiple times), block definition and cache (to store the results from running the blocks), worker information and the data units.

\subsection{Deployment}

Figure \ref{fig:flow} shows an example workflow using CrowdHub, where we define a crowdsourcing experiment following a between-subjects design to systematically evaluate task interface alternatives.

CrowdHub enables the entire task evaluation process, as shown in Figure \ref{fig:flow}. At \textit{design time}, requesters use the workflow editor to define the experimental design, which includes the tasks (\textit{Do} boxes) and the data flow (indicated by the arrows and the \textit{lambda} functions describing data aggregation and partitioning). Experimental groups can also be defined and associated to one or more tasks, denoted in the diagram using different colors. 
When deploying the experiment, the \textit{Workflow Manager} parses the workflow definition and creates the individual tasks in the target crowdsourcing platform with the associated data units and task design, relying on the Integration Layer to handle the selected platform. 
At \textit{run time}, the requester can specify the population management strategy and time sampling, if any, and the platform will leverage the Worker Manager and Scheduler modules to launch, pause and resume the tasks, and manage workers participation.
\section{Discussion}

Running controlled experiments in crowdsourcing environments is a challenging endeavor. Researchers must put special care in formulating the task and effectively communicating workers what they should perform \cite{DBLP:conf/chi/KitturCS08}. Unintended worker behavior could be observed if researchers fail at delivering the task; for example, poor instructions could lead workers to produce low-quality work or discourage them from participating in the first place \cite{DBLP:conf/hcomp/WuQ17}.
The inherent biases associated with experimental research and those present in the crowdsourcing platforms hinder the job of researchers. 
However, most of these biases are unknown to researchers approaching crowdsourcing platforms and could harm the experimental results, reducing the acceptance of crowdsourcing as an experimental method \cite{Crump2013,Paolacci2010RunningEO}.
For example, the underlying population demographics characterizes the workers that can take part in an experiment. If run uncontrolled, the sampled population may not be representative and include biases that hurt the experimental outcome \cite{DBLP:conf/chi/BarbosaC19}.

We showed specific instances of how running crowdsourcing experiments without coping strategies can impact the experimental design, assignment, and workers participating in the experiments.
Using task design evaluation, we distilled the challenges and quantified how it could change the outcomes of experiments.
However, these challenges are not only tied to task design evaluation, and in general, they play a role in the success of crowdsourcing experiments.
Qarout et al. \cite{Qarout2019PlatformRelatedFI} identified that worker performance could vary significantly across platforms, showing that workers in AMT performed the same task significantly faster than those on F8. 
This difference highlights the importance and impact of the underlying population demographics \cite{DBLP:conf/wsdm/DifallahFI18}.
Recurring workers naturally affects longitudinal studies. Over prolonged periods, the population of workers could refresh \cite{DBLP:conf/wsdm/DifallahFI18}; however, this still depends on the actual platform. Therefore, it is common for researchers to resort to employing custom mechanisms \cite{gadiraju2017improving} or leveraging the actual worker identifiers to limit participation \cite{Qarout2019PlatformRelatedFI}.
This forces researchers to fill the gaps by programming extensions to crowdsourcing platforms and make it possible to run controlled experiments, one of the core challenges highlighted by behavioral researchers \cite{Crump2013}.

The lack of native support from crowdsourcing platforms to deliver experimental protocols could potentially affect the validity and generalization of experimental results.
Researchers are forced to master advanced platform-dependent features or even extend their capabilities to implement coping strategies and bring control to crowdsourcing experiments.
This is because crowdsourcing platforms were built with micro-tasking and data collection tasks in mind where results are important.
However, the potential downside of manually extending crowdsourcing platforms is the learning curve that it incurs on researchers. This could lock researchers to specific platforms and discourage running experiments across multiple crowdsourcing vendors --- potentially threatening how well results generalize to other environments \cite{Qarout2019PlatformRelatedFI}.

The reliability of results, associated with choosing the right sample size, is a critical challenge in experimental design. The inherent cost of recruiting participants forces researchers to trade-off sample size, time, and budget in laboratory settings \cite{DBLP:conf/chi/KitturCS08}. For crowdsourcing environments, this is not necessarily the case, since it naturally represents easy access to a large pool of participants \cite{Crump2013}.
In laboratory settings, it is possible to resort to magic numbers and formulas to derive the sample size, but these rely on established recruitment criteria that can ensure a certain level of homogeneity --- in general, a homogenous population would require a small sample size. In crowdsourcing, however, it is not easy to screen participants, and sometimes researchers just accept whoever is willing to participate. This can bring a lot of variability into the results.
One way to address this issue is to rely on techniques used in adaptive or responsive survey design, i.e., stopping when we estimate that another round of crowdsourcing (e.g., data collection) will have a low probability of changing our current estimates \cite{RaoAdaptiveSurveys2008}. Another approach would be to run simulations inspired by k-fold cross validation \cite{StatisticalLearningBook}, which relies on specific data distribution (e.g., unimodal).

As crucial as controlling different aspects of task design, experimental protocol, and coping strategies to deal with the crowdsourcing environment is the fact that researchers must communicate these clearly to aid repeatable and reproducible crowdsourcing experiments \cite{Qarout2019PlatformRelatedFI,DBLP:conf/www/Paritosh12}.
In the context of systematic literature reviews, PRISMA \cite{Prisma} defines a thorough checklist that aids researchers in the preparation and reporting of robust systematic reviews.
For crowdsourcing researchers and practitioners, there is currently no concise guideline on what should be reported to facilitate reproducing results from crowdsourcing experiments, beyond those guidelines addressing concrete crowdsourcing processes \cite{DBLP:conf/naacl/LiuSBLLW16,DBLP:conf/chi/BarbosaC19,DBLP:conf/lrec/SabouBDS14,DBLP:conf/sigir/BlancoHHMPTT11}.
We find this an exciting direction for future work, and we are currently initiating our project on developing guidelines for reporting crowdsourcing experiments and encouraging repeatable and reproducible research.

We created CrowdHub to provide features that enable requesters to build crowdsourcing workflows, such as creating datasets for training machine learning models or designing and executing complex crowdsourcing experiments.

CrowdHub is not a monolithic system but rather a collection of components that interact with each other. 
Requesters can register new adapters with the \textit{Integration Layer}  to add support for new crowdsourcing platforms, therefore growing the list of vendors that CrowdHub provides out-of-the-box. This could be particularly useful for integrating private in-house platforms that suites the specific needs of task requesters.
The \textit{Workflow Manager} enables researchers and practitioners to add new \textit{blocks} to the set of available nodes for creating crowdsourcing workflows. 
With this feature, requesters can grow the scope of crowdsourcing workflows that can be created using CrowdHub. 
Therefore, new forms of computation could be added that run alongside crowd workers. An example is adding an ``ML block" that uses data units to train a machine learning classifier.

To create task interfaces, requesters can resort to the built-in set of UI elements that encapsulates the necessary code for rendering the interface on the underlying crowdsourcing platforms. CrowdHub's frontend application exposes these UI elements as visual boxes that requesters can drag and drop to arrange and configure the interface accordingly.
The current set consists of elements for rendering text, images, form text inputs, and inputs for multiple- and single-choice selection. 
We also incorporated the possibility of highlighting text and image elements.  This is useful, for example, to generate datasets for natural language processing and computer vision (e.g., question-answering and object detection datasets, respectively), or studying the impact of text highlighting in classification tasks \cite{ramirez2019}.
We plan to add support for actually coding the task interface, allowing requesters to use HTML, JavaScript, and CSS instead of using the current editor that offers draggable visual elements. 
This modality will give full flexibility to experienced requesters for designing task interfaces, and in this context, the current UI components will be available as special ``HTML tags".

% As we described previously, requesters can extend the set of UI elements to suite their needs, but they should also implement the corresponding ``renderers" and register these with the \textit{Integration Layer}. Concretely, each UI element has a corresponding ``renderer" (a JavaScript module) for each crowdsourcing platform adapter in the \textit{Integration Layer}. Of course, for new UI elements, requesters only need to implement the renderers for the crowdsourcing platform they target.
% %
% We plan to add support for actually coding the task interface, allowing requesters to use HTML, JavaScript, and CSS instead of using the current editor that offers draggable visual elements. 
% %
% This modality will give full flexibility to experienced requesters for designing task interfaces, and in this context, the current UI components will be available as special ``HTML tags".

% CrowdHub also offers features to support collaboration among requesters, particularly relevant in the context of running crowdsourcing experiments. Task requesters can share access to workflows with other collaborators, allowing other members to design and configure parts of the crowdsourcing workflow.
% %
% We also add support for generating URLs for read-only access to workflows. This feature aims to foster repeatability and reproducibility of results in crowdsourcing experiments --- crucial aspects for crowdsourcing research \cite{DBLP:conf/www/Paritosh12,Qarout2019PlatformRelatedFI}.

We presented a demo of CrowdHub \cite{CrowdHub2019} and received positive and constructive feedback from researchers in the human computation community. These discussions allowed us to arrive at the current design goals and set of features that constitute CrowdHub.
The current implementation offers all the features we described for the Workflow manager, and a subset of the functionality associated with the Worker Manager that enables eligibility control, allowing researchers to map experimental designs and control worker participation.
%. In particular, we implemented the eligibility control feature available in the Worker Manager module, which allows researchers to design between-subjects experiments and, overall, control worker participation in specific tasks. 
%The population management, as well as the Scheduler module, are still on process of implementation.
%
The system also supports collaboration between requesters and generating URLs for read-only access to workflows --- a feature that aims to foster repeatability and reproducibility of results in crowdsourcing experiments.
CrowdHub currently supports two crowdsourcing platforms: Figure Eight and Yandex Toloka. Support for Amazon Mechanical Turk is also in the roadmap.

\section{Conclusion}

In this paper, we draw from our experience and distilled the challenges and coping strategies to run controlled experiments in crowdsourcing environments.
Using the systematic evaluation of task design, we quantified the potential impact of uncontrolled crowdsourcing experiments and connected the challenges to those found in other crowdsourcing contexts.
Inspired by these lessons, and how frequently they occur in the literature, we designed and implemented CrowdHub, a system that extends crowdsourcing platforms and allows requesters to run controlled crowdsourcing projects. 
As part of our future work, we plan to run user studies to evaluate the extent to which CrowdHub supports researchers in running crowdsourcing experiments and practitioners in deploying crowdsourcing workflows.
CrowdHub is an open-source project, and we made available on Github the source code of the frontend\footnote{CrowdHub frontend: \url{https://github.com/TrentoCrowdAI/crowdhub-web}} and backend\footnote{CrowdHub backend: \url{https://github.com/TrentoCrowdAI/crowdhub-api}} layers.

% \section*{Acknowledgment}

% The preferred spelling of the word ``acknowledgment'' in America is without 
% an ``e'' after the ``g''. Avoid the stilted expression ``one of us (R. B. 
% G.) thanks $\ldots$''. Instead, try ``R. B. G. thanks$\ldots$''. Put sponsor 
% acknowledgments in the unnumbered footnote on the first page.

% \bibliographystyle{ACM-Reference-Format}
\bibliographystyle{IEEEtran}
\bibliography{references}

\begin{thebibliography}{10}
\providecommand{\url}[1]{#1}
\csname url@samestyle\endcsname
\providecommand{\newblock}{\relax}
\providecommand{\bibinfo}[2]{#2}
\providecommand{\BIBentrySTDinterwordspacing}{\spaceskip=0pt\relax}
\providecommand{\BIBentryALTinterwordstretchfactor}{4}
\providecommand{\BIBentryALTinterwordspacing}{\spaceskip=\fontdimen2\font plus
\BIBentryALTinterwordstretchfactor\fontdimen3\font minus
  \fontdimen4\font\relax}
\providecommand{\BIBforeignlanguage}[2]{{%
\expandafter\ifx\csname l@#1\endcsname\relax
\typeout{** WARNING: IEEEtran.bst: No hyphenation pattern has been}%
\typeout{** loaded for the language `#1'. Using the pattern for}%
\typeout{** the default language instead.}%
\else
\language=\csname l@#1\endcsname
\fi
#2}}
\providecommand{\BIBdecl}{\relax}
\BIBdecl

\bibitem{DBLP:journals/pvldb/JainSPW17}
\BIBentryALTinterwordspacing
A.~Jain, A.~D. Sarma, A.~G. Parameswaran, and J.~Widom, ``Understanding
  workers, developing effective tasks, and enhancing marketplace dynamics: {A}
  study of a large crowdsourcing marketplace,'' \emph{{PVLDB}}, vol.~10, no.~7,
  pp. 829--840, 2017. [Online]. Available:
  \url{http://www.vldb.org/pvldb/vol10/p829-dassarma.pdf}
\BIBentrySTDinterwordspacing

\bibitem{DBLP:journals/csur/DanielKCBA18}
\BIBentryALTinterwordspacing
F.~Daniel, P.~Kucherbaev, C.~Cappiello, B.~Benatallah, and M.~Allahbakhsh,
  ``Quality control in crowdsourcing: {A} survey of quality attributes,
  assessment techniques, and assurance actions,'' \emph{{ACM} Comput. Surv.},
  vol.~51, no.~1, pp. 7:1--7:40, 2018. [Online]. Available:
  \url{https://doi.org/10.1145/3148148}
\BIBentrySTDinterwordspacing

\bibitem{DBLP:conf/hcomp/WuQ17}
\BIBentryALTinterwordspacing
M.~Wu and A.~J. Quinn, ``Confusing the crowd: Task instruction quality on
  amazon mechanical turk,'' in \emph{{HCOMP} 2017}, 2017. [Online]. Available:
  \url{https://aaai.org/ocs/index.php/HCOMP/HCOMP17/paper/view/15943}
\BIBentrySTDinterwordspacing

\bibitem{DBLP:conf/cscw/KitturNBGSZLH13}
\BIBentryALTinterwordspacing
A.~Kittur, J.~V. Nickerson, M.~S. Bernstein, E.~Gerber, A.~D. Shaw,
  J.~Zimmerman, M.~Lease, and J.~J. Horton, ``The future of crowd work,'' in
  \emph{Computer Supported Cooperative Work, {CSCW} 2013, San Antonio, TX, USA,
  February 23-27, 2013}, 2013, pp. 1301--1318. [Online]. Available:
  \url{https://doi.org/10.1145/2441776.2441923}
\BIBentrySTDinterwordspacing

\bibitem{DBLP:conf/ht/GadirajuYB17}
\BIBentryALTinterwordspacing
U.~Gadiraju, J.~Yang, and A.~Bozzon, ``Clarity is a worthwhile quality: On the
  role of task clarity in microtask crowdsourcing,'' in \emph{Proceedings of
  the 28th {ACM} Conference on Hypertext and Social Media, {HT} 2017, Prague,
  Czech Republic, July 4-7, 2017}, 2017, pp. 5--14. [Online]. Available:
  \url{https://doi.org/10.1145/3078714.3078715}
\BIBentrySTDinterwordspacing

\bibitem{DBLP:conf/naacl/LiuSBLLW16}
\BIBentryALTinterwordspacing
A.~Liu, S.~Soderland, J.~Bragg, C.~H. Lin, X.~Ling, and D.~S. Weld, ``Effective
  crowd annotation for relation extraction,'' in \emph{{NAACL} {HLT} 2016, The
  2016 Conference of the North American Chapter of the Association for
  Computational Linguistics: Human Language Technologies, San Diego California,
  USA, June 12-17, 2016}, 2016, pp. 897--906. [Online]. Available:
  \url{http://aclweb.org/anthology/N/N16/N16-1104.pdf}
\BIBentrySTDinterwordspacing

\bibitem{DBLP:conf/ecis/SchulzeSGKS11}
\BIBentryALTinterwordspacing
T.~Schulze, S.~Seedorf, D.~Geiger, N.~Kaufmann, and M.~Schader, ``Exploring
  task properties in crowdsourcing - an empirical study on mechanical turk,''
  in \emph{19th European Conference on Information Systems, {ECIS} 2011,
  Helsinki, Finland, June 9-11, 2011}, 2011, p. 122. [Online]. Available:
  \url{http://aisel.aisnet.org/ecis2011/122}
\BIBentrySTDinterwordspacing

\bibitem{DBLP:conf/chi/SampathRI14}
\BIBentryALTinterwordspacing
H.~A. Sampath, R.~Rajeshuni, and B.~Indurkhya, ``Cognitively inspired task
  design to improve user performance on crowdsourcing platforms,'' in
  \emph{{CHI} Conference on Human Factors in Computing Systems, CHI'14,
  Toronto, ON, Canada - April 26 - May 01, 2014}, 2014, pp. 3665--3674.
  [Online]. Available: \url{https://doi.org/10.1145/2556288.2557155}
\BIBentrySTDinterwordspacing

\bibitem{Wilson2016WWW}
\BIBentryALTinterwordspacing
S.~Wilson, F.~Schaub, R.~Ramanath, N.~Sadeh, F.~Liu, N.~A. Smith, and F.~Liu,
  ``Crowdsourcing annotations for websites' privacy policies: Can it really
  work?'' in \emph{WWW 2018}, 2016. [Online]. Available:
  \url{https://doi.org/10.1145/2872427.2883035}
\BIBentrySTDinterwordspacing

\bibitem{ramirez2019}
J.~Ram\'{i}rez, M.~Baez, F.~Casati, and B.~Benatallah, ``Understanding the
  impact of text highlighting in crowdsourcing tasks.'' in \emph{Proceedings of
  the Seventh {AAAI} Conference on Human Computation and Crowdsourcing, {HCOMP}
  2019}, vol.~7.\hskip 1em plus 0.5em minus 0.4em\relax AAAI, October 2019, pp.
  144--152.

\bibitem{Whiting2019FairWC}
M.~E. Whiting, G.~Hugh, and M.~S. Bernstein, ``Fair work: Crowd work minimum
  wage with one line of code,'' in \emph{{HCOMP} 2019}, 2019.

\bibitem{DBLP:journals/sigkdd/MasonW09}
\BIBentryALTinterwordspacing
W.~A. Mason and D.~J. Watts, ``Financial incentives and the "performance of
  crowds",'' \emph{{SIGKDD} Explorations}, vol.~11, no.~2, pp. 100--108, 2009.
  [Online]. Available: \url{https://doi.org/10.1145/1809400.1809422}
\BIBentrySTDinterwordspacing

\bibitem{DBLP:conf/www/HoSSV15}
\BIBentryALTinterwordspacing
C.~Ho, A.~Slivkins, S.~Suri, and J.~W. Vaughan, ``Incentivizing high quality
  crowdwork,'' in \emph{Proceedings of the 24th International Conference on
  World Wide Web, {WWW} 2015, Florence, Italy, May 18-22, 2015}, 2015, pp.
  419--429. [Online]. Available: \url{https://doi.org/10.1145/2736277.2741102}
\BIBentrySTDinterwordspacing

\bibitem{maddalena2016crowdsourcing}
E.~Maddalena, M.~Basaldella, D.~De~Nart, D.~Degl'Innocenti, S.~Mizzaro, and
  G.~Demartini, ``Crowdsourcing relevance assessments: The unexpected benefits
  of limiting the time to judge,'' in \emph{Fourth AAAI Conference on Human
  Computation and Crowdsourcing}, 2016.

\bibitem{krishna2016embracing}
R.~A. Krishna, K.~Hata, S.~Chen, J.~Kravitz, D.~A. Shamma, L.~Fei-Fei, and
  M.~S. Bernstein, ``Embracing error to enable rapid crowdsourcing,'' in
  \emph{Proceedings of the 2016 CHI conference on human factors in computing
  systems}.\hskip 1em plus 0.5em minus 0.4em\relax ACM, 2016, pp. 3167--3179.

\bibitem{DBLP:journals/imwut/GadirajuCGD17}
\BIBentryALTinterwordspacing
U.~Gadiraju, A.~Checco, N.~Gupta, and G.~Demartini, ``Modus operandi of crowd
  workers: The invisible role of microtask work environments,'' \emph{{IMWUT}},
  vol.~1, no.~3, pp. 49:1--49:29, 2017. [Online]. Available:
  \url{https://doi.org/10.1145/3130914}
\BIBentrySTDinterwordspacing

\bibitem{Paolacci2010RunningEO}
G.~Paolacci, J.~Chandler, and P.~G. Ipeirotis, ``Running experiments on amazon
  mechanical turk,'' 2010.

\bibitem{Buhrmester2011AmazonsMT}
M.~D. Buhrmester, T.~N. Kwang, and S.~D. Gosling, ``Amazon's mechanical turk: A
  new source of inexpensive, yet high-quality, data?'' \emph{Perspectives on
  psychological science : a journal of the Association for Psychological
  Science}, vol. 6 1, pp. 3--5, 2011.

\bibitem{Schnoebelen2010UsingAM}
T.~Schnoebelen and V.~Kuperman, ``Using amazon mechanical turk for linguistic
  research,'' 2010.

\bibitem{DBLP:conf/icse/SunS16}
\BIBentryALTinterwordspacing
P.~Sun and K.~T. Stolee, ``Exploring crowd consistency in a mechanical turk
  survey,'' in \emph{Proceedings of the 3rd International Workshop on
  CrowdSourcing in Software Engineering, CSI-SE@ICSE 2016, Austin, Texas, USA,
  May 16, 2016}, 2016, pp. 8--14. [Online]. Available:
  \url{https://doi.org/10.1145/2897659.2897662}
\BIBentrySTDinterwordspacing

\bibitem{Crump2013}
\BIBentryALTinterwordspacing
M.~J.~C. Crump, J.~V. McDonnell, and T.~M. Gureckis, ``Evaluating amazon's
  mechanical turk as a tool for experimental behavioral research,'' \emph{PLOS
  ONE}, vol.~8, no.~3, pp. 1--18, 03 2013. [Online]. Available:
  \url{https://doi.org/10.1371/journal.pone.0057410}
\BIBentrySTDinterwordspacing

\bibitem{DBLP:conf/emnlp/SnowOJN08}
\BIBentryALTinterwordspacing
R.~Snow, B.~O'Connor, D.~Jurafsky, and A.~Y. Ng, ``Cheap and fast - but is it
  good? evaluating non-expert annotations for natural language tasks,'' in
  \emph{2008 Conference on Empirical Methods in Natural Language Processing,
  {EMNLP} 2008, Proceedings of the Conference, 25-27 October 2008, Honolulu,
  Hawaii, USA, {A} meeting of SIGDAT, a Special Interest Group of the {ACL}},
  2008, pp. 254--263. [Online]. Available:
  \url{http://www.aclweb.org/anthology/D08-1027}
\BIBentrySTDinterwordspacing

\bibitem{DBLP:conf/chi/KitturCS08}
\BIBentryALTinterwordspacing
A.~Kittur, E.~H. Chi, and B.~Suh, ``Crowdsourcing user studies with mechanical
  turk,'' in \emph{Proceedings of the 2008 Conference on Human Factors in
  Computing Systems, {CHI} 2008, 2008, Florence, Italy, April 5-10, 2008},
  2008, pp. 453--456. [Online]. Available:
  \url{https://doi.org/10.1145/1357054.1357127}
\BIBentrySTDinterwordspacing

\bibitem{DBLP:conf/uist/KitturSKK11}
\BIBentryALTinterwordspacing
A.~Kittur, B.~Smus, S.~Khamkar, and R.~E. Kraut, ``Crowdforge: crowdsourcing
  complex work,'' in \emph{Proceedings of the 24th Annual {ACM} Symposium on
  User Interface Software and Technology, Santa Barbara, CA, USA, October
  16-19, 2011}, 2011, pp. 43--52. [Online]. Available:
  \url{https://doi.org/10.1145/2047196.2047202}
\BIBentrySTDinterwordspacing

\bibitem{DBLP:conf/uist/AhmadBMK11}
\BIBentryALTinterwordspacing
S.~Ahmad, A.~Battle, Z.~Malkani, and S.~D. Kamvar, ``The jabberwocky
  programming environment for structured social computing,'' in
  \emph{Proceedings of the 24th Annual {ACM} Symposium on User Interface
  Software and Technology, Santa Barbara, CA, USA, October 16-19, 2011}, 2011,
  pp. 53--64. [Online]. Available:
  \url{https://doi.org/10.1145/2047196.2047203}
\BIBentrySTDinterwordspacing

\bibitem{DBLP:conf/cscw/KulkarniCH12}
\BIBentryALTinterwordspacing
A.~P. Kulkarni, M.~Can, and B.~Hartmann, ``Collaboratively crowdsourcing
  workflows with turkomatic,'' in \emph{{CSCW} '12 Computer Supported
  Cooperative Work, Seattle, WA, USA, February 11-15, 2012}, 2012, pp.
  1003--1012. [Online]. Available:
  \url{https://doi.org/10.1145/2145204.2145354}
\BIBentrySTDinterwordspacing

\bibitem{RamirezBMC2019}
\BIBentryALTinterwordspacing
J.~Ram{\'i}rez, M.~Baez, F.~Casati, and B.~Benatallah, ``Crowdsourced dataset
  to study the generation and impact of text highlighting in classification
  tasks,'' \emph{BMC Research Notes}, vol.~12, no.~1, p. 820, 2019. [Online].
  Available: \url{https://doi.org/10.1186/s13104-019-4858-z}
\BIBentrySTDinterwordspacing

\bibitem{DBLP:conf/chi/BarbosaC19}
\BIBentryALTinterwordspacing
N.~M. Barbosa and M.~Chen, ``Rehumanized crowdsourcing: {A} labeling framework
  addressing bias and ethics in machine learning,'' in \emph{Proceedings of the
  2019 {CHI} Conference on Human Factors in Computing Systems, {CHI} 2019,
  Glasgow, Scotland, UK, May 04-09, 2019}, 2019, p. 543. [Online]. Available:
  \url{https://doi.org/10.1145/3290605.3300773}
\BIBentrySTDinterwordspacing

\bibitem{DBLP:conf/lrec/BalahurSKZGHPB10}
\BIBentryALTinterwordspacing
A.~Balahur, R.~Steinberger, M.~A. Kabadjov, V.~Zavarella, E.~V. der Goot,
  M.~Halkia, B.~Pouliquen, and J.~Belyaeva, ``Sentiment analysis in the news,''
  in \emph{Proceedings of the International Conference on Language Resources
  and Evaluation, {LREC} 2010, 17-23 May 2010, Valletta, Malta}, 2010.
  [Online]. Available:
  \url{http://www.lrec-conf.org/proceedings/lrec2010/summaries/909.html}
\BIBentrySTDinterwordspacing

\bibitem{DBLP:conf/ht/ChengC13}
\BIBentryALTinterwordspacing
J.~Cheng and D.~Cosley, ``How annotation styles influence content and
  preferences,'' in \emph{24th {ACM} Conference on Hypertext and Social Media
  (part of ECRC), {HT} '13, Paris, France - May 02 - 04, 2013}, 2013, pp.
  214--218. [Online]. Available: \url{https://doi.org/10.1145/2481492.2481519}
\BIBentrySTDinterwordspacing

\bibitem{eickhoff2018cognitive}
C.~Eickhoff, ``Cognitive biases in crowdsourcing,'' in \emph{Proceedings of the
  Eleventh ACM International Conference on Web Search and Data Mining}.\hskip
  1em plus 0.5em minus 0.4em\relax ACM, 2018.

\bibitem{DBLP:conf/coling/NguyenTDGTMJ14}
\BIBentryALTinterwordspacing
D.~Nguyen, D.~Trieschnigg, A.~S. Dogru{\"{o}}z, R.~Gravel, M.~Theune, T.~Meder,
  and F.~de~Jong, ``Why gender and age prediction from tweets is hard: Lessons
  from a crowdsourcing experiment,'' in \emph{{COLING} 2014, 25th International
  Conference on Computational Linguistics, Proceedings of the Conference:
  Technical Papers, August 23-29, 2014, Dublin, Ireland}, 2014, pp. 1950--1961.
  [Online]. Available: \url{https://www.aclweb.org/anthology/C14-1184/}
\BIBentrySTDinterwordspacing

\bibitem{DBLP:conf/cscw/SenGGHLNRWH15}
\BIBentryALTinterwordspacing
S.~Sen, M.~E. Giesel, R.~Gold, B.~Hillmann, M.~Lesicko, S.~Naden, J.~Russell,
  Z.~K. Wang, and B.~J. Hecht, ``Turkers, scholars, "arafat" and "peace":
  Cultural communities and algorithmic gold standards,'' in \emph{Proceedings
  of the 18th {ACM} Conference on Computer Supported Cooperative Work {\&}
  Social Computing, {CSCW} 2015, Vancouver, BC, Canada, March 14 - 18, 2015},
  2015, pp. 826--838. [Online]. Available:
  \url{https://doi.org/10.1145/2675133.2675285}
\BIBentrySTDinterwordspacing

\bibitem{Qarout2019PlatformRelatedFI}
R.~K. Qarout, A.~Checco, G.~Demartini, and K.~Bontcheva, ``Platform-related
  factors in repeatability and reproducibility of crowdsourcing tasks,'' in
  \emph{{HCOMP} 2019}, 2019.

\bibitem{DBLP:conf/www/Paritosh12}
\BIBentryALTinterwordspacing
P.~Paritosh, ``Human computation must be reproducible,'' in \emph{Proceedings
  of the First International Workshop on Crowdsourcing Web Search, Lyon,
  France, April 17, 2012}, 2012, pp. 20--25. [Online]. Available:
  \url{http://ceur-ws.org/Vol-842/crowdsearch-paritosh.pdf}
\BIBentrySTDinterwordspacing

\bibitem{DBLP:conf/uist/BernsteinLMHAKCP10}
\BIBentryALTinterwordspacing
M.~S. Bernstein, G.~Little, R.~C. Miller, B.~Hartmann, M.~S. Ackerman, D.~R.
  Karger, D.~Crowell, and K.~Panovich, ``Soylent: a word processor with a crowd
  inside,'' in \emph{Proceedings of the 23rd Annual {ACM} Symposium on User
  Interface Software and Technology, New York, NY, USA, October 3-6, 2010},
  2010, pp. 313--322. [Online]. Available:
  \url{https://doi.org/10.1145/1866029.1866078}
\BIBentrySTDinterwordspacing

\bibitem{CrowdRev2018}
J.~Ram{\'{i}}rez, E.~Krivosheev, M.~B{\'{a}}ez, F.~Casati, and B.~Benatallah,
  ``Crowdrev: {A} platform for crowd-based screening of literature reviews,''
  in \emph{Collective Intelligence, {CI} 2018}, 2018.

\bibitem{DBLP:conf/criwg/CorreiaSPF18}
\BIBentryALTinterwordspacing
A.~Correia, D.~Schneider, H.~Paredes, and B.~Fonseca, ``Scicrowd: Towards a
  hybrid, crowd-computing system for supporting research groups in academic
  settings,'' in \emph{Collaboration and Technology - 24th International
  Conference, {CRIWG} 2018, Costa de Caparica, Portugal, September 5-7, 2018,
  Proceedings}, 2018, pp. 34--41. [Online]. Available:
  \url{https://doi.org/10.1007/978-3-319-99504-5\_4}
\BIBentrySTDinterwordspacing

\bibitem{DBLP:conf/sigmod/FranklinKKRX11}
\BIBentryALTinterwordspacing
M.~J. Franklin, D.~Kossmann, T.~Kraska, S.~Ramesh, and R.~Xin, ``Crowddb:
  answering queries with crowdsourcing,'' in \emph{Proceedings of the {ACM}
  {SIGMOD} International Conference on Management of Data, {SIGMOD} 2011,
  Athens, Greece, June 12-16, 2011}, 2011, pp. 61--72. [Online]. Available:
  \url{https://doi.org/10.1145/1989323.1989331}
\BIBentrySTDinterwordspacing

\bibitem{DBLP:conf/uist/LittleCGM10}
\BIBentryALTinterwordspacing
G.~Little, L.~B. Chilton, M.~Goldman, and R.~C. Miller, ``Turkit: human
  computation algorithms on mechanical turk,'' in \emph{Proceedings of the 23rd
  Annual {ACM} Symposium on User Interface Software and Technology, New York,
  NY, USA, October 3-6, 2010}, 2010, pp. 57--66. [Online]. Available:
  \url{https://doi.org/10.1145/1866029.1866040}
\BIBentrySTDinterwordspacing

\bibitem{DBLP:conf/socinfo/MinderB12}
\BIBentryALTinterwordspacing
P.~Minder and A.~Bernstein, ``Crowdlang: {A} programming language for the
  systematic exploration of human computation systems,'' in \emph{Social
  Informatics - 4th International Conference, SocInfo 2012, Lausanne,
  Switzerland, December 5-7, 2012. Proceedings}, 2012, pp. 124--137. [Online].
  Available: \url{https://doi.org/10.1007/978-3-642-35386-4\_10}
\BIBentrySTDinterwordspacing

\bibitem{DBLP:conf/oopsla/BarowyCBM12}
\BIBentryALTinterwordspacing
D.~W. Barowy, C.~Curtsinger, E.~D. Berger, and A.~McGregor, ``Automan: a
  platform for integrating human-based and digital computation,'' in
  \emph{Proceedings of the 27th Annual {ACM} {SIGPLAN} Conference on
  Object-Oriented Programming, Systems, Languages, and Applications, {OOPSLA}
  2012, part of {SPLASH} 2012, Tucson, AZ, USA, October 21-25, 2012}, 2012, pp.
  639--654. [Online]. Available: \url{https://doi.org/10.1145/2384616.2384663}
\BIBentrySTDinterwordspacing

\bibitem{DBLP:conf/cikm/ParameswaranPGPW12}
\BIBentryALTinterwordspacing
A.~G. Parameswaran, H.~Park, H.~Garcia{-}Molina, N.~Polyzotis, and J.~Widom,
  ``Deco: declarative crowdsourcing,'' in \emph{21st {ACM} International
  Conference on Information and Knowledge Management, CIKM'12, Maui, HI, USA,
  October 29 - November 02, 2012}, 2012, pp. 1203--1212. [Online]. Available:
  \url{https://doi.org/10.1145/2396761.2398421}
\BIBentrySTDinterwordspacing

\bibitem{DBLP:conf/cidr/DemartiniTKF13}
\BIBentryALTinterwordspacing
G.~Demartini, B.~Trushkowsky, T.~Kraska, and M.~J. Franklin, ``Crowdq:
  Crowdsourced query understanding,'' in \emph{{CIDR} 2013, Sixth Biennial
  Conference on Innovative Data Systems Research, Asilomar, CA, USA, January
  6-9, 2013, Online Proceedings}, 2013. [Online]. Available:
  \url{http://cidrdb.org/cidr2013/Papers/CIDR13\_Paper137.pdf}
\BIBentrySTDinterwordspacing

\bibitem{DBLP:journals/pvldb/MarcusWKMM11}
\BIBentryALTinterwordspacing
A.~Marcus, E.~Wu, D.~R. Karger, S.~Madden, and R.~C. Miller, ``Human-powered
  sorts and joins,'' \emph{{PVLDB}}, vol.~5, no.~1, pp. 13--24, 2011. [Online].
  Available:
  \url{http://www.vldb.org/pvldb/vol5/p013\_adammarcus\_vldb2012.pdf}
\BIBentrySTDinterwordspacing

\bibitem{DBLP:journals/pvldb/MorishimaSMAF12}
\BIBentryALTinterwordspacing
A.~Morishima, N.~Shinagawa, T.~Mitsuishi, H.~Aoki, and S.~Fukusumi,
  ``Cylog/crowd4u: {A} declarative platform for complex data-centric
  crowdsourcing,'' \emph{Proc. {VLDB} Endow.}, vol.~5, no.~12, pp. 1918--1921,
  2012. [Online]. Available:
  \url{http://vldb.org/pvldb/vol5/p1918\_atsuyukimorishima\_vldb2012.pdf}
\BIBentrySTDinterwordspacing

\bibitem{DBLP:conf/osdi/DeanG04}
\BIBentryALTinterwordspacing
J.~Dean and S.~Ghemawat, ``Mapreduce: Simplified data processing on large
  clusters,'' in \emph{6th Symposium on Operating System Design and
  Implementation {(OSDI} 2004), San Francisco, California, USA, December 6-8,
  2004}, 2004, pp. 137--150. [Online]. Available:
  \url{http://www.usenix.org/events/osdi04/tech/dean.html}
\BIBentrySTDinterwordspacing

\bibitem{DBLP:conf/hcomp/MaoCGPPZ12}
\BIBentryALTinterwordspacing
A.~Mao, Y.~Chen, K.~Z. Gajos, D.~C. Parkes, A.~D. Procaccia, and H.~Zhang,
  ``Turkserver: Enabling synchronous and longitudinal online experiments,'' in
  \emph{The 4th Human Computation Workshop, HCOMP@AAAI 2012, Toronto, Ontario,
  Canada, July 23, 2012}, 2012. [Online]. Available:
  \url{http://www.aaai.org/ocs/index.php/WS/AAAIW12/paper/view/5315}
\BIBentrySTDinterwordspacing

\bibitem{DBLP:conf/wsdm/DifallahFI18}
\BIBentryALTinterwordspacing
D.~E. Difallah, E.~Filatova, and P.~Ipeirotis, ``Demographics and dynamics of
  mechanical turk workers,'' in \emph{Proceedings of the Eleventh {ACM}
  International Conference on Web Search and Data Mining, {WSDM} 2018, Marina
  Del Rey, CA, USA, February 5-9, 2018}, 2018, pp. 135--143. [Online].
  Available: \url{https://doi.org/10.1145/3159652.3159661}
\BIBentrySTDinterwordspacing

\bibitem{gadiraju2017improving}
U.~Gadiraju and R.~Kawase, ``Improving reliability of crowdsourced results by
  detecting crowd workers with multiple identities,'' in \emph{International
  Conference on Web Engineering}.\hskip 1em plus 0.5em minus 0.4em\relax
  Springer, 2017, pp. 190--205.

\bibitem{RaoAdaptiveSurveys2008}
\BIBentryALTinterwordspacing
R.~S. Rao, M.~E. Glickman, and R.~J. Glynn, ``Stopping rules for surveys with
  multiple waves of nonrespondent follow-up,'' \emph{Statistics in Medicine},
  vol.~27, no.~12, pp. 2196--2213, 2008. [Online]. Available:
  \url{https://onlinelibrary.wiley.com/doi/abs/10.1002/sim.3063}
\BIBentrySTDinterwordspacing

\bibitem{StatisticalLearningBook}
G.~James, D.~Witten, T.~Hastie, and R.~Tibshirani, \emph{An Introduction to
  Statistical Learning: With Applications in R}.\hskip 1em plus 0.5em minus
  0.4em\relax Springer Publishing Company, Incorporated, 2013, pp. 181--186.

\bibitem{Prisma}
\BIBentryALTinterwordspacing
L.~Shamseer, D.~Moher, M.~Clarke, D.~Ghersi, A.~Liberati, M.~Petticrew,
  P.~Shekelle, and L.~A. Stewart, ``Preferred reporting items for systematic
  review and meta-analysis protocols (prisma-p) 2015: elaboration and
  explanation,'' \emph{BMJ}, vol. 349, 2015. [Online]. Available:
  \url{https://www.bmj.com/content/349/bmj.g7647}
\BIBentrySTDinterwordspacing

\bibitem{DBLP:conf/lrec/SabouBDS14}
\BIBentryALTinterwordspacing
M.~Sabou, K.~Bontcheva, L.~Derczynski, and A.~Scharl, ``Corpus annotation
  through crowdsourcing: Towards best practice guidelines,'' in
  \emph{Proceedings of the Ninth International Conference on Language Resources
  and Evaluation, {LREC} 2014, Reykjavik, Iceland, May 26-31, 2014}, 2014, pp.
  859--866. [Online]. Available:
  \url{http://www.lrec-conf.org/proceedings/lrec2014/summaries/497.html}
\BIBentrySTDinterwordspacing

\bibitem{DBLP:conf/sigir/BlancoHHMPTT11}
\BIBentryALTinterwordspacing
R.~Blanco, H.~Halpin, D.~M. Herzig, P.~Mika, J.~Pound, H.~S. Thompson, and
  D.~T. Tran, ``Repeatable and reliable search system evaluation using
  crowdsourcing,'' in \emph{Proceeding of the 34th International {ACM} {SIGIR}
  Conference on Research and Development in Information Retrieval, {SIGIR}
  2011, Beijing, China, July 25-29, 2011}, 2011, pp. 923--932. [Online].
  Available: \url{https://doi.org/10.1145/2009916.2010039}
\BIBentrySTDinterwordspacing

\bibitem{CrowdHub2019}
\BIBentryALTinterwordspacing
J.~Ram{\'{\i}}rez, S.~Degiacomi, D.~Zanella, M.~B{\'{a}}ez, F.~Casati, and
  B.~Benatallah, ``Crowdhub: Extending crowdsourcing platforms for the
  controlled evaluation of tasks designs,'' \emph{CoRR}, vol. abs/1909.02800,
  2019. [Online]. Available: \url{http://arxiv.org/abs/1909.02800}
\BIBentrySTDinterwordspacing

\end{thebibliography}

\end{document}